\documentclass[preprint,12pt]{elsarticle}

\usepackage{amsmath,amssymb,xspace,subfigure}
\usepackage{amsfonts}
\usepackage[usenames,dvipsnames]{xcolor}
\usepackage[]{graphicx}
\usepackage{url}
\usepackage{hyperref}
\usepackage{fullpage}
\usepackage{url}
\usepackage{float}
\usepackage{bm}
\usepackage{soul}
\usepackage{natbib}
\usepackage{physics}
\usepackage[ruled,vlined, noend]{algorithm2e}
\usepackage[noend]{algpseudocode}
\usepackage{booktabs}
\usepackage{siunitx}

\graphicspath{{images/}}

\journal{International Journal of Non-Linear Mechanics}

\begin{document}

\begin{frontmatter}

\title{Inverse Design of Planar Clamped-Free Elastic Rods from Noisy Data}

\author[a]{Dezhong Tong\corref{cor5}}
\author[a]{Zhuonan Hao}
\author[b]{Weicheng Huang\corref{cor5}}

\cortext[cor5]{Corresponding authors: tltl960308@g.ucla.edu (D.T.) and weicheng.huang@ncl.ac.uk (W.H.)}

\address[a]{Department of Mechanical and Aerospace Engineering, University of California, Los Angeles, \\ Los Angeles, California, 90095}
\address[b]{School of Engineering, Newcastle University, Stephenson Building, \\ Newcastle upon Tyne, NE1 7RU, UK}

\begin{abstract}
{

 Slender structures, such as rods, often exhibit large nonlinear geometrical deformations even under moderate external forces (e.g., gravity).
This characteristic results in a rich variety of morphological changes, making them appealing for engineering design and applications, such as soft robots, submarine cables, decorative knots, and more. Prior studies have demonstrated that the natural shape of a rod significantly influences its deformed geometry.
Consequently, the natural shape of the rod should be considered when manufacturing and designing rod-like structures. Here, we focus on an inverse problem: can we determine the natural shape of a suspended 2D planar rod so that it deforms into a desired target shape?
We begin by formulating a theoretical framework based on the statics of planar rod equilibrium that can compute the natural shape of a planar rod given its target shape.
Furthermore, we analyze the impact of uncertainties (e.g., noise in the data) on the accuracy of the theoretical framework. The results reveal the shortcomings of the theoretical framework in handling uncertainties in the inverse problem, a fact often overlooked in previous works.
To mitigate the influence of the uncertainties, we combine the statics of the planar rod with the adjoint method for parameter sensitivity analysis, constructing a learning framework that can efficiently explore the natural shape of the designed rod with enhanced robustness. This framework is validated numerically for its accuracy and robustness, offering valuable insights into the inverse design of soft structures for various applications, including soft robotics and animation of morphing structures.
}
\end{abstract}


\end{frontmatter}

\section{Introduction}
\label{sec:intro}

Rods, one-dimensional structures characterized by their long and thin geometry, are ubiquitous in the real world. These structures encompass a wide range of physical, biological, and manufactured phenomena, spanning from macroscopic examples such as hairs~\cite{kaufman2014adaptive}, tendrils~\cite{gerbode2012cucumber}, and cables~\cite{tong2024sim2real} to microscopic ones like DNA molecules~\cite{hogan1978transient} and carbon nanotubes~\cite{geblinger2008self}. Most rod-like structures are not naturally straight but feature an intricate, curved natural shape, leading to their structural richness, which benefits many engineering applications, including knots~\cite{choi2021imc, tong2023snap, tong2024fisherman}, soft robots~\cite{tong2023fully, hao2024bundling}, and surgery threads~\cite{pai2002strands, chentanez2009interactive}. Most prior works have explored the deformations and mechanics of these processes in a forward manner. However, designing the geometry and material parameters of deformable structures, given their deformed configurations and constraints, is sometimes more important for the design and fabrication of rod or beam-like engineering applications like 4D printing of composite structures~\cite{hyun2013foldable} and manufacturing of soft robots~\cite{felton2014method}. 

The inverse design of rod-like structures stems from various theories of elasticity, developed to describe the equilibrium and deformations of rods under finite displacements. 
In the 19th century, Kirchhoff and Clebsch proposed elastic theory for inextensible and unshearable rods~\cite{kirchhoff1859uber}, which was later expanded by the Cosserat brothers to include extensible and shearable elastic rods~\cite{cosserat1909theorie}.
Numerous research efforts have since focused on developing analytical and numerical approaches to solving the Kirchhoff rod equations for diverse applications, such as studying DNA molecules~\cite{neukirch2004extracting, bouchiat2000elastic} and plant growth~\cite{goriely2006mechanics, gerbode2012cucumber}, as well as robotics applications involving deformable structures like wire management~\cite{tong2024sim2real, tong2021automated} and sheet folding~\cite{choi2024learning}. 
All of those works emphasize the significant influence of natural curvature on the deformed configurations of rods and analyze the deformation of rod-like structures in a forward manner. Those works on the forward process underpin the inverse design approaches discussed in this article.

As the name suggests, the inverse problem involves determining the natural shape or material properties of a rod-like structure from its deformed equilibrium and constraints. 
The focus here is on inverse design, where the material parameters are known and the objective is to discover the natural shape of the rod~\cite{bertails2018inverse, qin2022bottom}, rather than inverse measurements, which aims to identify material properties like bending stiffness~\cite{derouet2013inverse, beck1998inverse, fachinotti2008finite}.
To tackle inverse design problems, various analytical and numerical approaches have been developed by different communities, including applied mathematics, mechanics, and computer graphics~\cite{turco2017tools}.
Those approaches can be classified into two main streams. The first stream formulates the inverse problem as elastica problems based on different elastic theories to generate general solutions. Examples include the inverse design for manipulating flexural waves on thin elastic planes using the Kirchhoff-Love equations~\cite{capers2023inverse}, the inverse design of morphing structures with tapered elastica~\cite{liu2020tapered}, and the inverse exploration of a rod's natural configuration using the Kirchhoff rod models~\cite{bertails2018inverse}.
The second stream involves numerical modeling approaches such as mass-spring systems and finite element methods (FEM), often assisted by nonlinear optimization algorithms to handle inverse problems in more general, nonlinear systems. For instance, Chen et al.\cite{chen2014asymptotic} combined the asymptotic numerical method with FEM to tackle the inverse elastic shape design during 3D printing. Topology optimization methods are used to compute the inverse design of the underwater metasurfaces~\cite{he2023inverse}, insulators~\cite{nanthakumar2019inverse}, nano structures~\cite{molesky2018inverse}, and mechanical springs~\cite{bluhm2023inverse}. The most relevant work to this article is Ref~\cite{qin2022bottom}, which utilized a bottom-up optimization method to explore the inverse solution of a clamped-free rod expressed by a mass-spring model.

However, few prior works consider the influence of uncertainties during the inverse design process. %
Noise is always present when measuring and modeling deformed elastic shapes, significantly affecting the accuracy of inverse design approaches.
Addressing these modeling and measurement uncertainties often necessitates a cumbersome process of data processing and model modification~\cite{turco2017tools}. 
Consequently, most of the above-stated works assume clean experimental data or just stall in the simulations, neglecting the impact of noise on their outcomes.

As data science advances, data-driven approaches show great potential in tackling the inverse design of engineering problems. 
For example, physics-informed neural networks (PINNs)\cite{cuomo2022scientific} and neural ordinary differential equations (neural ODEs)\cite{chen2018neural} have demonstrated significant promise in encoding the physics of a system within a neural network.
Machine learning communities introduce physical laws as constraints to guide the training of models to meet specific design requirements. 
These data-driven approaches have become popular in inverse design problems due to their adjustable regularization and better robustness to uncertainties compared to traditional inverse design approaches~\cite{yang2021b}. 
For instance, Lu et al.~\cite{lu2021physics} successfully implemented PINNs to solve a series of inverse problems in solid and fluid mechanics. 
However, considering the gray-box properties of these physics-informed neural networks, the accuracy of solutions remains a bottleneck for data-driven inverse design approaches.

In this study, we leverage the theoretical foundation of traditional inverse design approaches and the data-driven framework from the machine learning community to propose an innovative solution for exploring the inverse design of elastic structures. 
Specifically, we focus on the inverse design problem of a planar rod: given the material properties and the targeted shape of the rod under gravity, our objective is to determine the natural shape of the rod. 
We represent the natural shape of the rod using a surrogate model and combine this with elastic theory to formulate the inverse design solutions. Our approach aims to achieve not only sufficient accuracy but also enhanced robustness against uncertainties in the system. Moreover, we release all our code for the proposed scheme as open-source software.\footnote{See \url{https://github.com/DezhongT/Inverse_Design_2D_Rods}.}

The following article is structured as follows. Sec.~\ref{sec:problem_description} discusses the problem that needs to be solved in detail. Sec.~\ref{sec:theory_base} provides the theoretical foundation, including the equilibrium of planar rods and the theoretical solution of the inverse problem. Sec.~\ref{sec:inverse_design_solution} discusses the harmfulness of uncertainties in the modeling and measurements for the theoretical solution and details our proposed method for tackling the inverse design problem with uncertainties. Sec.~\ref{sec:experiments} showcases experimental results. Finally, Sec.~\ref{sec:conclusion} concludes the paper and shows potential future work. 

\begin{figure}[h!]
    \centering
    \includegraphics[width=1\textwidth]{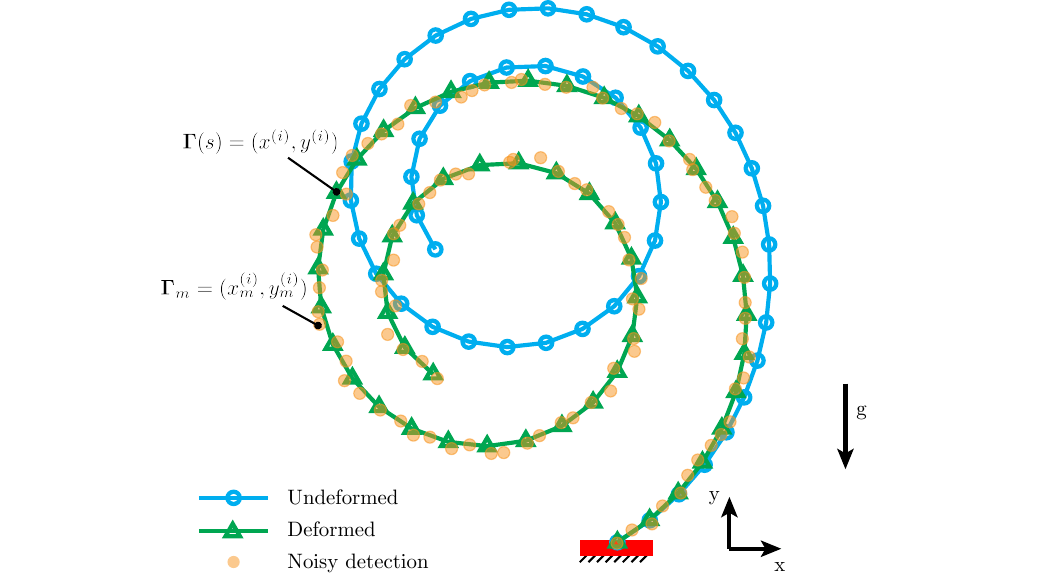}
    \caption{Diagram illustrating the deformation process of an elastic rod with a clamped-free configuration under the influence of gravitational acceleration $g$. The detection and segmentation of the structure are likely to be performed using a finite number of pixel points, as yellow dotted points, with white noise present in the measurement process. 
    }
    \label{fig:overview}
\end{figure}

\section{Problem description}
\label{sec:problem_description}
This study focuses on the inverse design problem of a two-dimensional clamped-free elastic rod subjected to gravitational force (as shown in Fig.~\ref{fig:overview}). 
Specifically, the goal is to reconstruct the undeformed, natural shape of the rod from its measured deformed configuration $\boldsymbol{\Gamma}_m$ and known material properties.
Notably, the detection of the deformed configuration is typically performed using digital sensors, such as cameras, which introduce quantization noise during the measurement process—a factor often ignored in prior works.
This noise complicates the inverse design process of soft rods.

The primary objective of this paper is to propose a robust and efficient scheme to extract the natural shape of the elastic rod accurately.
The proposed method aims to ensure that the rod's reconstructed deformed configuration $\boldsymbol{\Gamma}$ under gravity closely matches the noisy measured data $\boldsymbol{\Gamma}_m$. The robust and accurate inverse design scheme is critical for applications that rely on precise deformation modeling and structural analysis.

\section{Background: Equilibrium of a 2D Planar Rod}
\label{sec:theory_base}
Before delving into the inverse design problem, let us straighten out the fundamental principles governing the intrinsic statics of a planar rod subject to gravity. 

In this work, we focus on a two-dimensional scenario involving an inextensible and unsharable rod of length $S$. This rod is depicted by a center line $\boldsymbol{\Gamma} \in \mathbb{R}^2$, along with a rotation angle $\theta \in \mathbb R$, both parameterized by arc length $s \in [0, S]$. To simplify the interpretation of the rod's geometry, we define a two-dimensional Cartesian coordinate and designate the end located at $s = 0$ as the origin and the negative direction of the gravity as the $\hat{\mathbf y}$-axis. Hereafter, all vectors with $\hat{(\cdot)}$ are unit vectors. At given location $s$, the vector $\boldsymbol{\Gamma}(s)$ denotes the 2D position $(x, y)$ of the center line, while the rotation $\theta(s)$ encodes the bending angles attached to the rod's cross-section.
Note that we simplify notations of variables that are functions of $s$, such as $\boldsymbol{\Gamma(s)}$, to $\boldsymbol{\Gamma}$ throughout this manuscript's equations.
Given the rod is assumed to be inextensible and unshearable, the following equation holds true:
\begin{equation}
\label{eq:geometry}
\forall s \in [0, S] \quad \boldsymbol{\Gamma}' = [\cos\theta, \sin\theta] ,
\end{equation}
where $()'$ stands for the first derivative with respect to $s$. We further assume that the rod is fixed at $s = 0$, indicating this end is clamped. Meanwhile, the other end at $s = S$ is free.

\subsection{Forward mechanics model}
\label{sec:rod_statics}
\begin{figure}[h!]
    \centering
    \includegraphics[width=\textwidth]{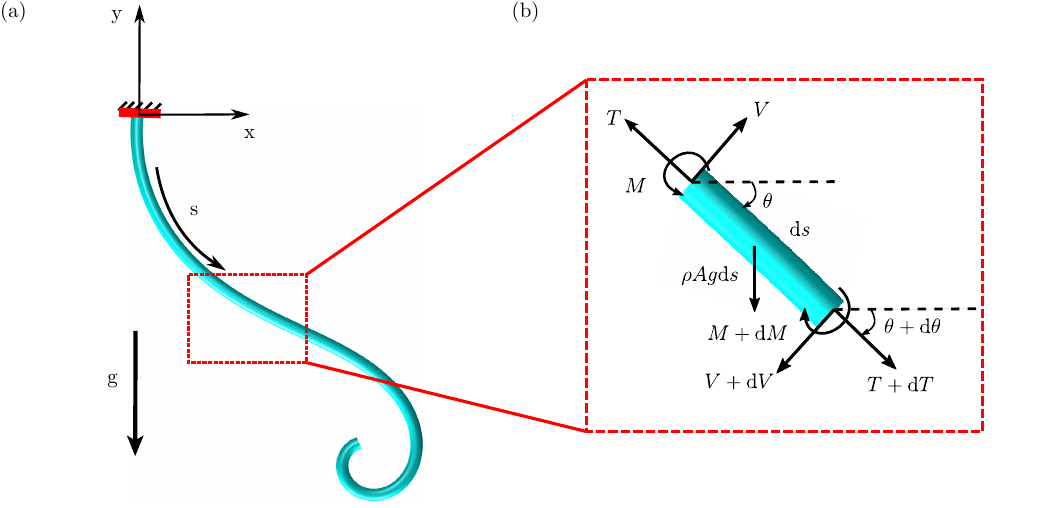}
    \caption{
    The schematic of a planar rod under gravity and the forces acting on its element.
    }
    \label{fig:planar_rod}
\end{figure}

To establish the equilibrium equations of the planar rod under gravity, we start with an element $\textrm{d}s$ of the rod. The element and the acting forces are shown in Fig.~\ref{fig:planar_rod}(b). A tension force $T(s)$ is acting in the direction of the tangent at position $s$, and the shear force, defined as the force perpendicular to a tension force, is denoted as $V(s)$. In the Cartesian coordinate, $\theta(s)$ is defined by the angle between the tangent and the horizontal axis $\hat{\mathbf x}$. In this manuscript, we assume the rod is only subjected to the known external force, such as gravity, which can be expressed by $\rho A g \textrm{d}s$, where $\rho$ is the volumetric mass density, $A$ is the cross-sectional area, and $g$ is the gravitational acceleration. Note that density $\rho$ and cross-sectional $A$ area are treated as constants during the analysis.
For the element in equilibrium, the sum of forces must be zero, which can derive the following equations:

\begin{equation}
\label{eq:force_balance}
\begin{aligned}
\dv{}{s} \left( T \cos\theta + V\sin\theta \right) &= 0 \\
\dv{}{s} \left( T \sin\theta - V\cos\theta\right) &= \rho A g 
\end{aligned}
\end{equation}
Meanwhile, the sum of the moments must be zero as well, which leads to:
\begin{equation}
\label{eq:moment_balance}
\begin{aligned}
V = \dv{M}{s}
\end{aligned}
\end{equation}
Let $E$ be the Young modulus of the rod, and $I$ be the second moment of area of the rod's cross-section. Static equations are complemented by a constitutive law that characterizes the rod's elastic bending behavior:
\begin{equation}
\label{eq:constitituve_law}
\forall s \in [0, S] \quad  M = EI (\kappa - \kappa_0) = EI (\dv{\theta}{s} - \kappa_0),
\end{equation}
where $EI$ is the bending stiffness of the planar rod, $\kappa(s)  = \dv{\theta(s)}{s}\in R$ is the curvature of the planar rod, and $\kappa_0(s) \in R$ encompasses the natural strains of the rod, which characterizes natural curvature, namely, the shape of the rod would assume in the absence of external forces, which may not be straight. Given modeling the homogeneous material, the stiffness $EI$ is considered constant. Conversely, the natural strain $\kappa_0(s)$ may vary spatially to encapsulate a diverse range of natural shapes.
Next, combining Eqs.~\ref{eq:moment_balance} and~\ref{eq:constitituve_law}, we obtain:
\begin{equation}
\label{eq:V_theta}
V = EI (\dv[2]{\theta}{s} - \dv{\kappa_0}{s}),
\end{equation}
and we introduce new variables:
\begin{equation}
\label{eq:new_variables}
\begin{aligned}
\bar{M} &= \frac{M}{\rho A g}\\
\bar{V} &= \frac{V}{\rho A g}\\
\bar{T} &= \frac{T}{\rho A g}.
\end{aligned}
\end{equation}
By substituting Eqs.~\ref{eq:V_theta} and~\ref{eq:new_variables} to Eq.~\ref{eq:force_balance}, we can obtain:
\begin{equation}
\label{eq:force_balance_new}
\begin{aligned}
\dv{}{s} \left( \bar T \cos\theta + \frac{1}{\eta} (\dv[2]{\theta}{s} -\dv{\kappa_0}{s}) \sin\theta\right) &= 0 \\
\dv{}{s} \left( \bar T \sin\theta - \frac{1}{\eta} (\dv[2]{\theta}{s} -\dv{\kappa_0}{s}) \cos\theta\right) &= 1, 
\end{aligned}
\end{equation}
where $\eta = \frac{\rho A g}{EI}$ characterizes the bending deformations under the external forces (gravity). Since a system of first-order differential equations (ODE) can be created from the higher-order ODEs, the system of the first ODEs can be constructed by combing Eqs.~\ref{eq:geometry},~\ref{eq:V_theta},~\ref{eq:new_variables}, and~\ref{eq:force_balance_new}:
\begin{equation}
\label{eq:ODE}
\begin{aligned}
 \mathbf{q}  &= [\theta, \bar M, \bar V,  \bar T, x, y]^T\\
\mathcal{R}(\mathbf q) &=  \dv{\mathbf q}{s} - f(\mathbf q) = 0\\
f(\mathbf q, \kappa_0) &= \begin{bmatrix}
    \eta q_2 + \kappa_0 \\
    q_3 \\
    -\cos q_1 + \eta q_4 (q_2 + \frac{1}{\eta} \kappa_0)\\
     \sin q_1 - \eta q_3 (q_2 + \frac{1}{\eta} \kappa_0)\\
     \cos q_1\\
     \sin q_1
\end{bmatrix}
\end{aligned},
\end{equation}
where $q_i$ means the $i-$th element of $\mathbf q(s)$.
Note that all quantities, including material properties $\eta$ and natural strains $\kappa_0$, are known when solving the forward pass of the rod's statics. 

Moving forward, we list the boundary conditions for the ODEs.
First, the clamped boundary at location $s = 0$ is:
\begin{equation}
\label{eq:clamped_BC}
\textrm{Clamped: }\left\{
\begin{array}{l}
\begin{aligned}
\boldsymbol{\Gamma}(s = 0) &= \boldsymbol{\Gamma}_0,  \quad \textrm{Enforced clamped position}\\
\theta(s = 0) &= \theta_0, \quad \textrm{Enforced clamped rotation}
\end{aligned}
\end{array}
\right.
\end{equation}
where $\boldsymbol{\Gamma}_s$ and $\theta_s$ are the given values.
Second, the free boundary at location $s = S$ is:
\begin{equation}
\label{eq:free_BC}
\textrm{Free: }\left\{
\begin{array}{l}
\begin{aligned}
\bar M(s = S) &= 0,  \quad \textrm{No external torque at the free end}\\
\bar T(s = S) &= \bar V(s= S) = 0, \quad \textrm{No external force at the free end}
\end{aligned}
\end{array}
\right.
\end{equation}
Combining the rod's governing equation stated in Eq.~\ref{eq:ODE} and boundary conditions in Eqs.~\ref{eq:clamped_BC} and~\ref{eq:free_BC}, we can get the equilibrium configuration of a planar rod in arbitrary configuration, which is depicted by $\kappa_0(s)$.

\subsection{Inverse design formulation}
\label{subsec:elastica}

As our interest here is the inverse design for the 2D planar rod, and, from the input target curve $\boldsymbol{\Gamma}(s)$, we aim at finding its natural curvature $\kappa_0(s)$ so that $\boldsymbol{\Gamma}(s)$ coincides with the center line of the planar rod at equilibrium under gravity.
Starting from the rod's governing equation stated in Eq.~\ref{eq:ODE}, we can outline the following equation:
\begin{equation}
\label{eq:formula_T}
\bar T = \frac{\theta^{'''} - \kappa_0^{''} + \eta \cos \theta}{\eta \theta'},
\end{equation}
We can calculate $\bar T'(s)$ based on Eq.~\ref{eq:formula_T}:
\begin{equation}
\label{eq:derivative_T}
\begin{aligned}
\bar T' &= \frac{-\theta'\kappa_0^{'''} + \theta^{''}\kappa_0^{''} + \theta^{''''} \theta' - \theta^{'''}\theta^{''} - \eta \theta'^2(\sin\theta + \cos\theta)}{\eta \theta'}.
\end{aligned}
\end{equation}
Then, substituting Eq.~\ref{eq:derivative_T} to Eq.~\ref{eq:ODE}, we can obtain the following equation:
\begin{equation}
\label{eq:inverse_theory}
\begin{aligned}
&\theta'\kappa_0^{'''} - \theta^{''}\kappa_0^{''} + \theta'^3\kappa_0' - \theta^{''''}\theta' + \theta^{'''}\theta^{''} + \theta^{''}(\eta \cos \theta - \theta'^3) -2\eta \theta'^2\sin \theta = 0.
\end{aligned}
\end{equation}
When the curve $\mathbf{\Gamma}(s)$ is determined, all $\theta(s)$ and its derivatives are known. Therein, Eq.~\ref{eq:inverse_theory} is an ordinary differential equation for variable $\kappa_0'(s)$. Once the boundary conditions are determined, we can obtain the theoretical solution of the inverse design problem by solving Eq.~\ref{eq:inverse_theory}.
The boundary conditions for the Eq.~\ref{eq:inverse_theory}:
\begin{equation}
\label{eq:inverse_BC}
\begin{aligned}
\textrm{Free: }
\kappa_0'(s=S) &= \theta''(s=S),\\
\textrm{Clamped: }
\kappa_0'(s=0) &= \theta''(s=0) - \eta\cos (\theta(s=0)).
\end{aligned}
\end{equation}

However, we can find that the solution of Eq.~\ref{eq:inverse_theory} is $\kappa'(s)$. Therein, we need to do the integration to obtain $\kappa(s)$. To calculate the natural strains of the planar rod, we need to give the initial boundary condition for $\kappa(s)$:
\begin{equation}
\label{eq:inverse_IC}
\begin{aligned}
\textrm{Free: }
\kappa_0(s=S) &= \theta'(s=S).
\end{aligned}
\end{equation}
Combining Eqs.~\ref{eq:inverse_theory},~\ref{eq:inverse_BC} and~\ref{eq:inverse_IC}, we can solve the elastica of the inverse problem.

\section{Inverse Design from Noisy Data}
\label{sec:inverse_design_solution}

In Sec.~\ref{subsec:elastica}, we present the elastica for the inverse design of a planar rod. 
However, our analysis above does not account for the influence of uncertainties in this engineering problem. Given the theoretical framework is based on the expression of $\theta(s)$, which is usually from the measured target shape $\boldsymbol{\Gamma}_m$. 
Therein, the uncertainties in this problem primarily come from two aspects: the uncertainties from the modeling and the uncertainties that exist in the measurements. 
To address the gap brought by the uncertainties, we first investigate the impact of uncertainties on the formulated theoretical framework. Then, we explore strategies to mitigate the adverse effects of uncertainties on the inverse design process.

\subsection{Uncertatinties in the modeling}
\begin{figure}[h!]
    \centering
    \includegraphics[width=1.0\textwidth]{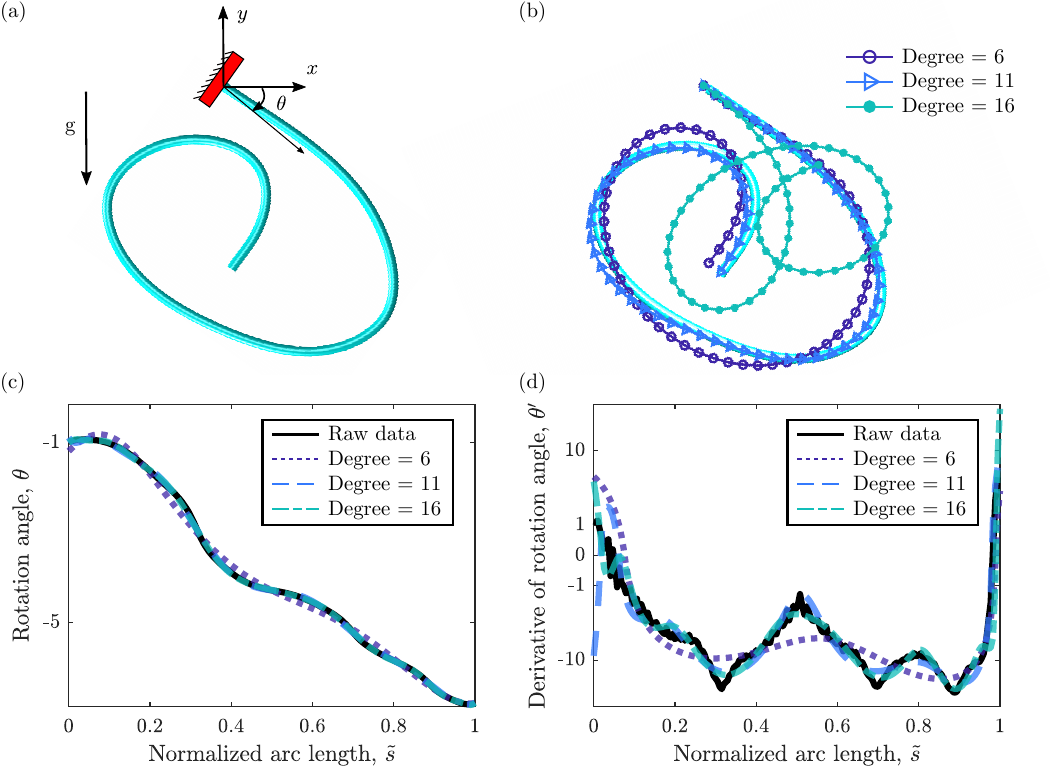}
    \caption{
    Influence of the chosen surrogate model for expression $\theta(s)$. (a) The designed target shape -- letter ``A''. (b) The comparison of the deformed shapes computed from the natural shape solved by different fitted models. (c) The comparison between raw data and fitted data from the polynomial regression model with degrees 6, 11, and 16 for $\theta(s)$. (d) The comparison between raw data and fitted data for $\theta'(s)$. Note that the normalized arc length $\Tilde{s} = s/S$.}
    \label{fig:model_influence}
\end{figure}

When computing the theoretical solution derived from Eq.~\ref{eq:inverse_theory}, the analysis is established on $\theta(s)$ obtaining by the combination of $\boldsymbol{\Gamma}(s)$ and Eq.~\ref{eq:geometry}. Given the solution's accuracy is closely tied to the high-order derivatives of $\theta(s)$, which are highly sensitive to the selection of the surrogate model used to describe $\theta(s)$, both overfitting and underfitting of the surrogate model can lead to significant deviations in the solution.

In Fig.~\ref{fig:model_influence}, we illustrate the influence of the chosen model on the theoretical solutions with an example. Here, we aim to design the natural shape of a rod so that it forms the letter ``A'' under gravity, as shown in Fig.~\ref{fig:model_influence}(a).
The letter ``A'' is extracted from a handwritten note using a digital camera, so that the sensor measurement noise exists in the image.
For simplicity, we use a polynomial regression model to represent $\theta(s)$ calculated from the pattern of ``A''.
When the polynomial degree of the fitting model is low (e.g., 6), the model is underfitted, leading to a noticeable difference between the fitted model and the raw data of $\theta(s)$, as shown in Fig.~\ref{fig:model_influence} (c).
Increasing the polynomial degree (e.g., 16) eliminates such difference but can result in overfitting. As shown in Fig.~\ref{fig:model_influence}(d), excessively high fitting degrees cause significant deviations in $\theta'(s)$.
We compare the accuracy of theoretical solutions from different fitted models of $\theta(s)$ as shown in Fig.~\ref{fig:model_influence}(b). 
The stated theoretical framework does not have mechanisms for such a model's uncertainties.
Thus, when parameterizing $\theta(s)$, managing the uncertainties associated with the model selection is crucial.

\subsection{Influence of the noisy data}
\label{sec:noisy_influence}
\begin{figure}[h!]
    \centering
    \includegraphics[width=1.0\textwidth]{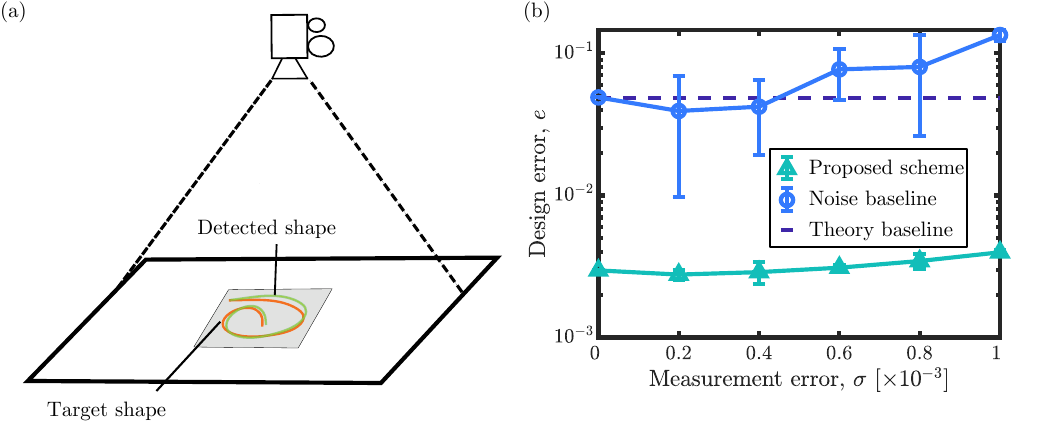}
    \caption{
      Influence of measurement noise on inverse design solutions. (a) Schematic of measuring the target shape, the letter "A," using a sensor (e.g., camera). (b) Impact of measurement error $\sigma$ on the accuracy of different approaches. The theory baseline represents the result obtained from the theoretical framework without added noise; the noise baseline represents the result obtained from the theoretical framework with added noise; the proposed scheme is our proposed optimal method (the learning framework detailed in Sec.~\ref{subsec:learning_framework}) designed to be robust against uncertainties. 
    }
    \label{fig:noise_influence}
\end{figure}

In real-world engineering applications, sensors such as cameras are commonly employed to measure a rod's deformed configuration $\boldsymbol{\Gamma}(s)$. 
However, it is crucial to acknowledge that sensor usage makes uncertainty and noise inevitable. In addition to modeling uncertainties, measurement noise becomes a significant source of uncertainty in this problem.
Therein, the measured curve dataset $ \boldsymbol{\Gamma}_m =\{(x_m^{(i)}, \; y_m^{(i)})\}_{i=1}^N$ for the inverse design problem can be:
\begin{equation}
\label{eq:noise_gamma}
\begin{aligned}
x_m^{(i)} &= x^{(i)} + \epsilon_x^{(i)},\\
y_m^{(i)} &= y^{(i)} + \epsilon_y^{(i)},\\
\end{aligned}
\end{equation}
where $\epsilon_x^{(i)}$ and $\epsilon_y^{(i)}$ are independent Gaussian noises with zero mean. We also presume that the fidelity of the sensor is predetermined, i.e., the standard deviations of $\epsilon_x^{(i)}$ and $\epsilon_y^{(i)}$ are known to be $\sigma_x^{(i)}$ and $\sigma_y^{(i)}$. In Fig.~\ref{fig:noise_influence}(b), we show the influence of Gaussian noise on the accuracy of the theoretical solutions obtained from Eq.~\ref{eq:inverse_theory}.
Since Eq.~\ref{eq:inverse_theory} is heavily dependent on the precision of high-order derivatives of $\theta(s)$, which are particularly sensitive to noise in measured data $\boldsymbol{\Gamma}_m$, the uncertainty existing in this system significantly diminishes the accuracy of the theoretical solutions. A robust scheme for the inverse design problem is imperative to compensate for the detrimental effects of the above-stated uncertainties.

\subsection{Learning framework -- the optimal inverse design solution}
\label{subsec:learning_framework}

In the theoretical framework, we need to model $\theta(s)$ and evaluate its high-order derivatives to obtain the designed parameters $\kappa_0(s)$. 
However, due to diffusion of measurement noise affecting the higher order derivatives of $\theta(s)$ and the uncertainty in model selection, directly employing this framework may lead to considerable deviations in determining the design parameter $\kappa_0(s)$.
To address this challenge, we propose a novel forward framework aimed at directly evaluating the measured target shape $\boldsymbol{\Gamma}_m$ with improved robustness against the uncertainties.
This approach enables us to avoid directly modeling $\theta(s)$ and mitigate the impact of measurement errors, thus enhancing the accuracy of solving for the design parameter $\kappa_0(s)$ within the framework.

The proposed forward framework is inspired by the physically informed neural networks (PINN)~\cite{cuomo2022scientific}, which has proven its efficacy in solving inverse design problems in many different physical scenarios. Here, we treat 
the design parameter, natural curvature $\kappa_0$ as an unknown, then the ODE in Eq.~\ref{eq:ODE} becomes:
\begin{equation}
\begin{aligned}
\mathcal{R}(\mathbf q, \kappa_0) = \dv[]{\mathbf q}{s} - f(\mathbf q, \kappa_0) = 0,
\end{aligned}
\end{equation}
where $\mathbf q = \mathbf q(s)$ is the forward solution of the deformed rod. Here, we start by representing $\kappa_0$ with a surrogate model $\kappa_0(s, \boldsymbol{\phi})$, where $\boldsymbol{\phi}$ is the vector of parameters in the surrogate model. Then, we can rewrite Eq.~\ref{eq:ODE} as:
\begin{equation}
\label{eq:ODE_with_param}
\begin{aligned}
\mathcal{R}(\mathbf q, \kappa_0(s, \boldsymbol{\phi})) = 0.
\end{aligned}
\end{equation}
By solving Eq.~\ref{eq:ODE_with_param} with an explicit Runge-Kutta method~\cite{kierzenka2001bvp}, we can obtain the expression of $x(s, \boldsymbol{\phi})$ and $y(s, \boldsymbol{\phi})$ directly. 
Then, the likelihood can be calculated as:
\begin{equation}
\label{eq:MLE}
\begin{aligned}
P(\boldsymbol{\Gamma}_m|\boldsymbol{\phi}) &= P(\boldsymbol{\Gamma}_m^x|\boldsymbol{\phi})P(\boldsymbol{\Gamma}_m^y|\boldsymbol{\phi}), \\
P(\boldsymbol{\Gamma}_m^x|\boldsymbol{\phi}) &= \prod_{i=1}^N \frac{1}{\sqrt{2\pi {\sigma_x^{(i)}}^2}} \exp(- \frac{(x_m^{(i)} - x^{(i)})^2}{2{\sigma_x^{(i)}}^2}), \\
P(\boldsymbol{\Gamma}_m^y|\boldsymbol{\phi}) &= \prod_{i=1}^N \frac{1}{\sqrt{2\pi {\sigma_y^{(i)}}^2}} \exp(- \frac{(y_m^{(i)} - y^{(i)})^2}{2{\sigma_y^{(i)}}^2}).
\end{aligned}
\end{equation}
We take the logarithm on both sides of Eq.~\ref{eq:MLE} to get the loss function:
\begin{equation}
\label{eq:MSE}
\begin{aligned}
L(\phi) = -\log P(\boldsymbol{\Gamma}_m|\boldsymbol{\phi}) \sim \frac{1}{N} \sum_{i=1}^N((x_m^{(i)} - x^{(i)})^2 +  (y_m^{(i)} - y^{(i)})^2),
\end{aligned}
\end{equation}
which is just the classical mean squared error (MSE) loss. In our designed training loop, MSE loss should be minimized to obtain the optimal parameters $\boldsymbol{\phi}$ for depicting the surrogate model $\kappa_0$.

\subsection{Adjoint method -- back differentitation}
In Eq.~\ref{eq:MSE}, we can find the relationship of the loss function to the parameters $\boldsymbol{\phi}$ of the surrogate model of $\kappa_0$ is implicit since $x^{(i)}$ and $y^{(i)}$ are obtained by solving Eq.~\ref{eq:ODE}.
Therefore, the main technical difficulty in the training loop is exploring the reverse-mode differentiation.
Considering the memory cost and computational speed, using numerical schemes like finite element difference to compute the gradient $\grad_{\phi} L$ is inappropriate here. 
Inspired by the backward differentiation in Neural ODE~\cite{chen2018neural}, we compute the gradients using the adjoint sensitivity method~\cite{pontryagin2018mathematical}. 
This approach computes gradients by solving a second, augmented ODE backward in time and is applicable to the training loop. This approach scales linearly with problem size, has low memory cost, and explicitly controls numerical error.

To formulate the adjoint method, we first rewrite the loss to an optimization object:
\begin{equation}
\label{eq:loss2}
\begin{aligned}
L(\mathbf q, \boldsymbol{\phi}) &\simeq \int g(\boldsymbol{\phi}, s) \textrm{d}s = \mathbf q_d^T \mathbb Q \mathbf q_d \textrm{d}s,\\
\end{aligned}
\end{equation}
with
\begin{equation}
\mathbf q_d =  \mathbf q - [0, 0, 0, 0, x_m, y_m]^T,\\
\end{equation}
and 
\begin{equation}
\mathbb Q =  \begin{bmatrix}
0 & 0 & 0 & 0 & 0 & 0 \\
0 & 0 & 0 & 0 & 0 & 0 \\
0 & 0 & 0 & 0 & 0 & 0 \\
0 & 0 & 0 & 0 & 0 & 0 \\
0 & 0 & 0 & 0 & 1 & 0 \\
0 & 0 & 0 & 0 & 0 & 1
\end{bmatrix},
\end{equation}
where the loss defined here is equally effective to the loss in Eq.~\ref{eq:MSE}. Therein, we can formulate a constrained problem:
\begin{equation}
\begin{aligned}
\min_\phi \quad & L(\mathbf q, \boldsymbol{\phi}) \notag \\
\text{s.t.} \quad & \dv[]{\mathbf q}{s} = \mathbf f(\mathbf q, s, \boldsymbol{\phi}). 
\end{aligned}
\end{equation}
To solve the constrained optimization, we can formulate a Lagrangian:
\begin{equation}
\mathcal{L} = \int (g(\mathbf q, \boldsymbol{\phi}) + \boldsymbol{\lambda}^T (\mathbf f - \dv[]{\mathbf q}{s}))\textrm{d}s, \; \mathrm{ with } \;
\boldsymbol{\lambda} \in \mathbb R^6
\end{equation}
We employ the differentiation on two sides:
\begin{equation}
\label{eq:opt}
\begin{aligned}
\dv[]{\mathcal{L}}{\boldsymbol{\phi}} = \int (\frac{\partial g}{\partial \boldsymbol{\phi}} + \boldsymbol{\lambda}^T \frac{\partial \boldsymbol{f}}{\partial \boldsymbol{\phi}} +  (\frac{\partial g}{\partial \mathbf q} + 
\boldsymbol{\lambda}^T\frac{\partial \mathbf f}{\partial \mathbf q} + \dv[]{\boldsymbol{\lambda}}{s}) \dv[]{\mathbf q}{\boldsymbol{\phi}}) \textrm{d}s + \boldsymbol{\lambda}(0)^T \dv[]{\mathbf q(0)}{\boldsymbol{\phi}}(0) -  \boldsymbol{\lambda}(S)^T \dv[]{\mathbf q(S)}{\phi}.
\end{aligned}
\end{equation}
Here, we choose $\boldsymbol{\lambda}(s)$, which can be solved by:
\begin{equation}
\label{eq:param_ode}
\begin{aligned}
\dv[]{\boldsymbol{\lambda}}{s} = -\frac{\partial g}{\partial \mathbf q} - 
\boldsymbol{\lambda}^T\frac{\partial \mathbf f}{\partial \mathbf q},
\end{aligned}
\end{equation}
with the specified boundary conditions as follows:

\begin{equation}
\label{eq:param_BC}
\begin{aligned}
\textrm{Clamped: } \quad
\boldsymbol{\lambda}_i(s = 0) = 0 \quad \text{for}\; i=2,3,4,\\
\textrm{Free: } \quad
\boldsymbol{\lambda}_i(s = S) = 0 \quad \text{for}\; i=1,5,6.\\
\end{aligned}
\end{equation}
Then, plugging the solution of $\boldsymbol{\lambda}(s)$ to Eq.~\ref{eq:opt}, the gradient of loss can be obtained:
\begin{equation}
\label{eq:grad}
\begin{aligned}
\grad_{\phi} L = \dv[]{\mathcal{L}}{\boldsymbol{\phi}} = \int (\frac{\partial g}{\partial \boldsymbol{\phi}} + \boldsymbol{\lambda}^T \frac{\partial f}{\partial \boldsymbol{\phi}})\textrm{d}s = \int \boldsymbol{\lambda}^T \frac{\partial f}{\partial \boldsymbol{\phi}} \textrm{d}s
\end{aligned}
\end{equation}
With the help of the computed gradient $\grad_\phi L$, we utilize Adam optimizer -- one of the most classical machine learning optimizers -- to train the surrogate model. The full algorithm is shown in Algo.~\ref{alg:inverse_design}.

\begin{algorithm}
\SetAlgoLined
\LinesNumbered
\DontPrintSemicolon
\KwIn{$\boldsymbol{\Gamma}_m, \eta, S$}
\KwOut{$\kappa_0$}
\SetKwProg{Fn}{Func}{:}{}
\SetKwFunction{Adam}{Adam}
$\alpha \gets$ learning rate of Adam optimizer \;
$\beta_1, \beta_2 \gets$ moment estimate decay rate \;
$\epsilon \gets$ a small constant \;
\SetKwFunction{InverseDesign}{InverseDesign}
{
\Fn{\InverseDesign{$\boldsymbol{\Gamma}_m, \eta, S$}}
{
$\theta^{(0)} \gets$ initialized with Eq.~\ref{eq:geometry} \;
$\kappa_0^{(0)} \gets$ initialized with Eqs.~\ref{eq:inverse_theory},~\ref{eq:inverse_BC} and~\ref{eq:inverse_IC}\;
$\boldsymbol{\phi}^{(0)} \gets$ initialized by regressing $\kappa_0^{(0)}$ \;
$L \gets$ initialize with a large constant \;
$i \gets$ 0 \;
\While{\textup{$i <$ max\underline{\phantom{X}}iter}}{
$\kappa_0(s) \gets$ computed with the surrogate model $\kappa_0(s; \boldsymbol{\phi})$ \;
$\mathbf q(s) \gets$ solved by Eqs.~\ref{eq:ODE},~\ref{eq:clamped_BC} and~\ref{eq:free_BC} \;
$L \gets$ loss computed by Eq.~\ref{eq:loss2}\;
\If{$L < \textup{small threshold}$}
{
    break \;
}
$\boldsymbol{\lambda}(s) \gets$ solved by the augmented ODE(Eq.~\ref{eq:param_ode}) and corresponding BCs(Eq.~\ref{eq:param_BC}) \;
$\grad_{\boldsymbol{\phi}} L \gets$ computed with Eq.~\ref{eq:grad} \;
$\Delta \boldsymbol{\phi} \gets$ \Adam{$\alpha, \beta_1, \beta_2, \epsilon, \boldsymbol{\phi}, \grad_{\boldsymbol{\phi}} L$} \;
$\boldsymbol{\phi}^{(i+1)} \gets \boldsymbol{\phi}^{(i)} - \Delta \boldsymbol{\phi}$\;
$i \gets i+1$\;
}
$\boldsymbol{\phi}^* \gets \boldsymbol{\phi}^{(i)}$  \;
$\kappa_0 \gets$  computed by the surrogate model $\kappa_0(s; \boldsymbol{\phi}^*)$ \;
\textbf{return} $\kappa_0$ \;

}
}
\caption{Inverse Design Process}
\label{alg:inverse_design}
\end{algorithm}

\section{Numerical Validation}
\label{sec:experiments}
In this section, we present comprehensive quantitative results to evaluate the performance of our proposed learning framework for the inverse design of planar rods. First, we compare the performance of the theoretical framework and the learning framework in processing artificially generated noisy curves.
Subsequently, we explore an engineering application of the inverse design process: can we design planar rods with varying stiffness to form a target pattern designed by users? We use a camera to detect and extract the pattern drawn on paper, which is then used as input for our proposed scheme. The results further demonstrate the efficacy of our method.

\subsection{Comparison between the theoretical and learning frameworks}
\begin{figure}[h!]
    \centering
    \includegraphics[width=\textwidth]{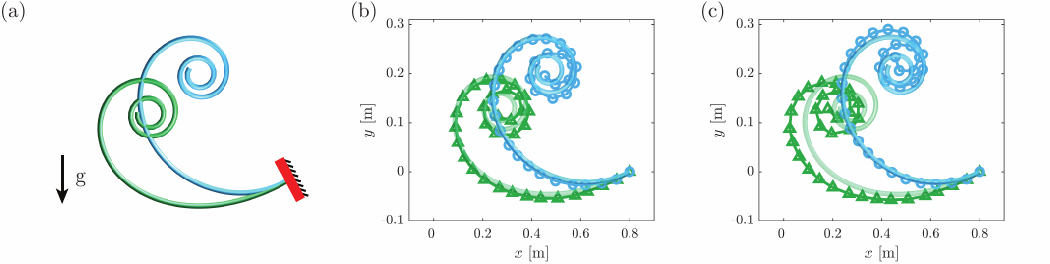}
    \caption{A randomly generated curve used to validate the effect of varying noise levels, characterized by the standard deviation $\sigma = 0.002$ (unit: \si{m}). The deformed shapes (green) are computed from the natural shape (blue) obtained by the fitted model. (a) Ground truth. (b) Predicted shape through optimal inverse design from noisy data. (c) Predicted shapes through inverse solver from noisy data. 
    }
    \label{fig:noise}
\end{figure}

Here, we generated multiple artificial target curves to evaluate the performance of the theoretical and learning frameworks. Generally, the target pattern $\boldsymbol{\Gamma}$, which is the deformed configuration of a planar rod, should be smooth due to the rod's elasticity. Since we aim to mimic a real engineering scenario, measurement noise should be introduced to these smooth curves. The protocol for preparing the noisy measured dataset of target shapes is as follows.
First, we define the rotation angle $\theta(s)$ using a polynomial function with coefficients randomly generated within a specific range, e.g., $[-10, 10]$. We then recover the deformed curve $\boldsymbol{\Gamma}(s)$ from the randomly generated $\theta(s)$. Next, we sample points from the deformed curve. To each sampling point, we add artificial Gaussian noise with a specific standard deviation $\sigma$ to mimic measurement noise. The resulting sampled discrete noisy dataset $\boldsymbol{\Gamma}_m$ is then used as input to compute the natural shape of the designed planar rod, as illustrated in Fig.~\ref{fig:noise}.

Once the measured dataset $\boldsymbol{\Gamma}_m$ is obtained, we utilize both the theoretical and learning frameworks to solve the inverse design problem. Hereafter, all quantities related to the theoretical framework are denoted with the superscript ``theory,'' and quantities related to our learning framework are denoted with ``opt.''
For the theoretical framework, we need to construct a model to parameterize $\theta^\textrm{theory}(s)$ from $\boldsymbol{\Gamma}_m$. This parameterized $\theta^\textrm{theory}(s)$ is then used in Eq.~\ref{eq:inverse_theory} to determine the natural curvature $\kappa^\textrm{theory}_0$ of the rod. As for the learning framework, we can compute the natural curvature $\kappa^\textrm{opt}_0$ with Algo.~\ref{alg:inverse_design}.
By substituting natural curvature $\kappa_0$ into Eq.~\ref{eq:ODE}, we can solve the rod's deformed configuration $\boldsymbol{\Gamma}$, corresponding to the different frameworks. 
The performance of the different frameworks is evaluated by comparing the average difference $e = \underset{s \in [0, S]}{\text{mean}} ||\boldsymbol{\Gamma}(s) - \boldsymbol{\Gamma}_m||/S$. A smaller difference $e$ value indicates better performance.

In Fig.~\ref{fig:noise}, we illustrate the impact of noisy measured data on the results of different frameworks. We can find that the theoretical solution is sensitive to noise, as shown in Fig.~\ref{fig:noise}(c), while our proposed learning framework is robust against the noise, accurately determining the natural shape of the planar rod so that it can deform to the prescribed pattern exactly, as shown in Fig.~\ref{fig:noise}(b).

\begin{figure}[h!]
    \centering
    \includegraphics[width=1.0\textwidth]{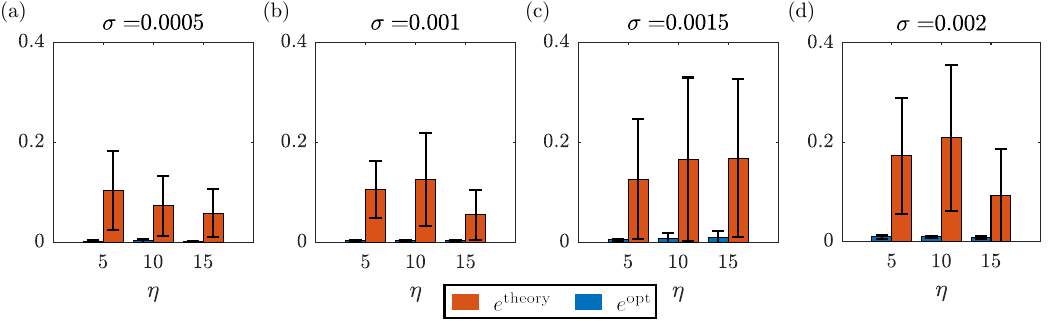}
    \caption{Evaluation of the inverse design accuracy for various target patterns with different measurement noise and material properties. $e^\textrm{theory}$ denotes the difference between the deformed configuration derived from the natural shape solved with the theoretical framework and the measured target shape, while $e^\textrm{opt}$ denotes the difference between the deformed configuration derived from the natural shape solved with the learning framework and the measured target shape. For each combination of the standard deviation of the measurement noise $\sigma$ (unit: \si{m}) and the rod's material properties $\eta$, 10 target patterns are randomly generated to evaluate the inverse design framework's accuracy.
    }
    \label{fig:error}
\end{figure}

To comprehensively evaluate the performance of the two frameworks, we randomly generate 120 target shapes and assess how each framework handles planar rods with varying stiffness. The differences between the predicted deformed configurations and the noise-measured data are given in Fig.~\ref{fig:error}. We can find $e^\textrm{opt}$ is much smaller compared to the $e^\textrm{theory}$. The difference between $e^\textrm{opt}$ and $e^\textrm{theory}$ becomes more significant when the standard deviation of the noise is larger, indicating that our learning framework performs better in robustly reconstructing the natural shape of a planar rod from noisy data. 
We also evaluate the performance of both frameworks for planar rods with different stiffness. The learning framework consistently achieves high performance across various planar rods with different stiffness demonstrating that our proposed learning framework can achieve robust and accurate inverse design results for rods made from various materials, whereas the theoretical framework generally performs poorly in nearly all cases.

\begin{figure}[h!]
    \centering
    \includegraphics[width=1.0\textwidth]{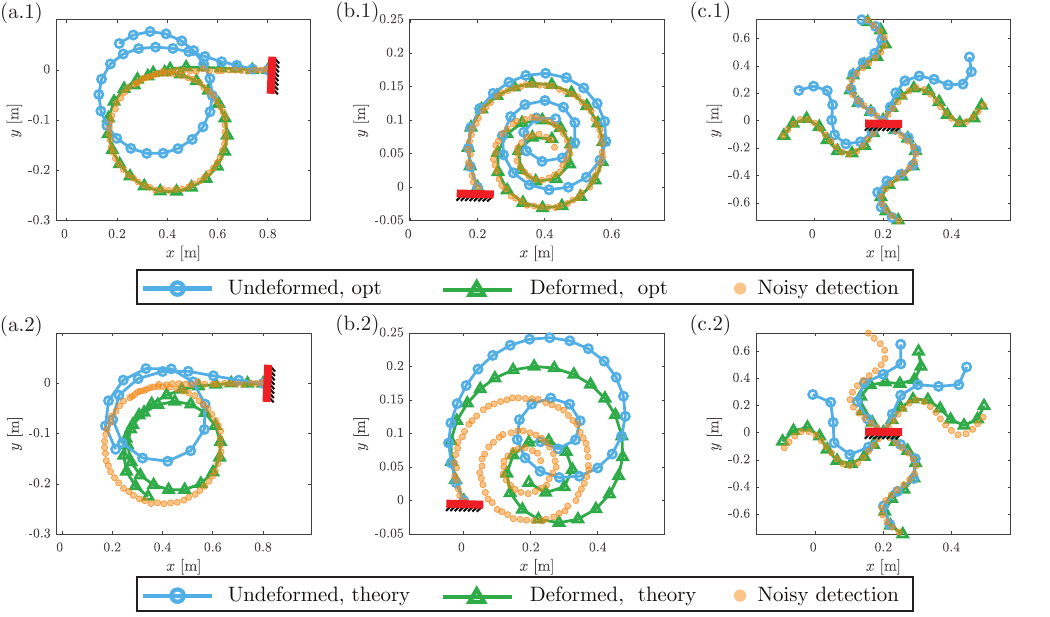}
    \caption{Three sample cases demonstrating variations in geometric patterns: (a) circle ($\eta = 15, \sigma = 0.002$), (b) spiral ($\eta = 15, \sigma = 0.002$), and (c) cross composed of intertwined sine curves ($\eta = 5, \sigma = 0.002$).
    }
    \label{fig:pattern}
\end{figure}

The performance of the learning framework for the inverse design is also illustrated with three canonical cases: a circle, a spiral, and a cross composed of sinusoidal curves in Fig.~\ref{fig:pattern}.

\subsection{Application -- Mimic human writing}
\begin{figure}[h!]
    \centering
    \includegraphics[width=1\textwidth]{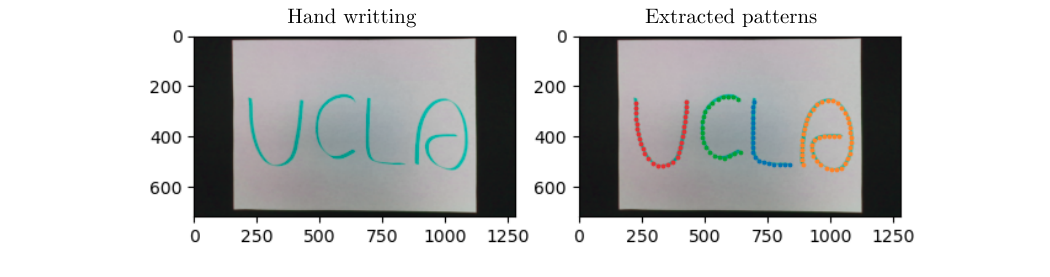}
    \caption{Handwritten letters and the corresponding extracted
    discretized patterns.
    }
    \label{fig:CV}
\end{figure}

\begin{figure}[h!]
    \centering
    \includegraphics[width=1.0\textwidth]{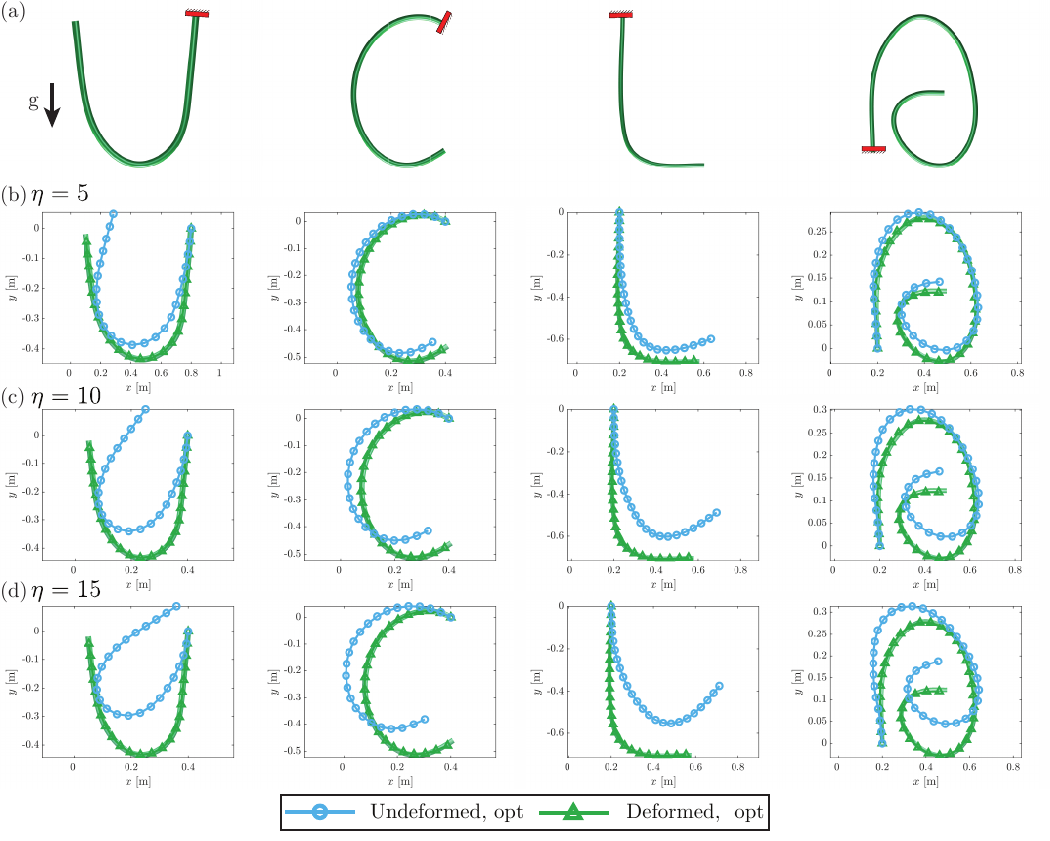}
    \caption{
    Inverse design of rods with different material properties to formulate the hand-written letters ``U'', ``C'', ``L'', and ``A'' with the learning framework.
    }
    \label{fig:UCLA}
\end{figure}

In this section, we demonstrate the ability to construct target shapes derived from human writing with the learning framework. Computer vision has become a powerful tool in various engineering fields, including pattern recognition and rapid measurement. With recent advancements in detecting deformable linear objects~\cite{choi2023mbest}, we can now use a camera (Realsense D435i) to extract patterns from human writing in real time, enabling us to easily obtain the desired target shapes. Here, we showcase the pattern acquisition process for the four letters ``U'', ``C'', ``L'', and ``A'', as shown in Fig.~\ref{fig:CV}. 

Once the discretized patterns are obtained, we can design the target shape and boundary conditions for the inverse design problem, as shown in Fig.~\ref{fig:UCLA}(a). Here, we employ our proposed learning framework to execute inverse design for planar rods with different stiffness.
First, we need to select a surrogate model to express $\kappa_0(s)$. This surrogate model can be any parameterized model, e.g., polynomial regression model, neural network, etc. Here, we use a polynomial regression model. This model has a single controlling parameter: the fitting degree number $n$. We select a relatively high degree number $n=14$ to model $\kappa_0(s)$ to prevent the underfitting issue.
As shown in Fig.~\ref{fig:UCLA}, our proposed learning framework can generalize to different handwriting patterns and material properties. 
The solutions generated by our method closely match the target shapes with minimal discrepancy, demonstrating its robustness against measurement noise and its effective regularization on the surrogate model.

\section{Conclusions}
\label{sec:conclusion}
In this article, we combine the elastic theory and machine learning algorithms to propose an efficient and robust learning framework for the inverse design of elastic rods from noisy measurement data. 
Inspired by physically informed neural networks, we design a forward structure based on the elastic theory of rods. We then define a loss function to represent the inverse design objective, enabling us to use reverse differentiation to optimize the relevant parameters -- the natural shape of the rod.
With comprehensive numerical validations, our proposed learning framework is proven to be valid in exploring the natural shape of the rod to minimize the designed loss function so that excellent agreements can be found between the target patterns and designed deformed rods.
The learning framework provides an effective scheme to solve inverse design problems of soft structures with imperfect measurements. The framework can provide valuable insights into real-world manufacturing of flexible structures such as soft robot design.

\bibliographystyle{elsarticle-num}
\bibliography{references}

\end{document}